\documentclass[twocolumn, superscriptaddress, prb, showpacs]{revtex4-1}

\usepackage{graphicx}
\usepackage{amsmath}
\usepackage[normalem]{ulem}
\usepackage{bbm}
\usepackage{mathbbol}

\newcommand{\equ}[1]{Eq.~(\ref{eq:#1})}

\newcommand{\eqs}[2]{Eqs.~(\ref{eq:#1}) and (\ref{eq:#2})}
\newcommand{\myhat}[1]{\hat{#1}}

\def\tr{{\rm tr}}
\def\Tr{{\rm Tr}}
\def\P{\myhat{P}}
\def\Q{\myhat{Q}}
\def\H{\myhat{H}}
\def\ham{\myhat{\cal H}}

\def\One{\myhat{\mathbbm{1}}}
\def\PP{\myhat{\mathbbm{P}}}
\def\QQ{\myhat{\mathbbm{Q}}}
\def\HH{\mathbbm{H}}
\def\AA{\mathbbm{A}}
\def\BB{\mathbbm{B}}
\def\CC{\mathbbm{C}}
\def\FF{\mathbbm{F}}
\def\MM{\mathbbm{M}}
\def\OO{\mathbb{\Omega}}
\def\LL{\mathbb{\Lambda}}

\def\Fab{F_{\alpha\beta}}
\def\Gab{G_{\alpha\beta}}
\def\Hab{H_{\alpha\beta}}

\def\lo{0}
\def\hi{1}

\def\nwann{J}
\def\nbands{{\cal J}}
\def\rop{\hat{r}}
\def\fermi{\varepsilon_F}

\def\k{{\bf k}}

\def\b{{\bf b}}
\def\M{{\bf M}}
\def\Mab{M_{\alpha\beta}}
\def\Im{{\rm Im}}
\def\Re{{\rm Re}}
\def\R{{\bf R}}
\def\r{{\bf r}}
\def\0{{\bf 0}}
\def\enk{\varepsilon_{n\k}}

\def\unk{u_{n\k}}

\def\ww{{{\rm W}}}
\def\hh{{{\rm H}}}
\def\Nq{{\cal N}}

\def\Hk{\H_\k}
\def\bra#1{\langle#1|}
\def\ket#1{|#1\rangle}
\def\braket#1#2#3{\langle#1|#2|#3\rangle}
\def\me#1#2{\langle#1|#2\rangle}

\def\dk{\partial_\k}
\def\da{\partial_\alpha}

\def\nn{\nonumber\\}
\def\beq{\begin{equation}}
\def\eeq{\end{equation}}

\def\eikr{e^{i\k\cdot\R}}
\def\emikr{e^{-i\k\cdot\R}}
\def\wt#1{\widetilde{#1}}

\begin{document}

\title{Wannier-based calculation of the orbital magnetization 
in crystals}

\author{M.~G.~Lopez}
\affiliation{Wake Forest University, Department of Physics,
Winston-Salem, NC 27109, USA}

\author{David Vanderbilt}
\affiliation{Department of Physics and Astronomy, Rutgers, The State 
University of New Jersey, Piscataway, New Jersey 08854, USA}

\author{T.~Thonhauser}
\email[E-mail: ]{thonhauser@wfu.edu}
\affiliation{Wake Forest University, Department of Physics,
Winston-Salem, NC 27109, USA}

\author{Ivo Souza}
\affiliation{Centro de F\'{\i}sica de Materiales (CSIC) and DIPC,
Universidad del Pa\'\i s Vasco, 20018 San Sebasti\'an, Spain}
\affiliation{Ikerbasque, Basque Foundation for Science, 48011 Bilbao,
Spain}

\date{\today}

\begin{abstract}
  We present a first-principles scheme that allows the orbital
  magnetization of a magnetic crystal to be evaluated accurately and
  efficiently even in the presence of complex Fermi surfaces.  Starting
  from an initial electronic-structure calculation with a coarse {\it ab
  initio} $\k$-point mesh, maximally localized Wannier functions are
  constructed and used to interpolate the necessary $\k$-space
  quantities on a fine mesh, in parallel to a previously-developed
  formalism for the anomalous Hall conductivity [X.~Wang, J.~Yates,
  I.~Souza, and D.~Vanderbilt, Phys.~Rev.~B {\bf 74}, 195118 (2006)].
  We formulate our new approach in a manifestly gauge-invariant manner,
  expressing the orbital magnetization in terms of traces over matrices
  in Wannier space. Since only a few (e.g., of the order of 20) Wannier
  functions are typically needed to describe the occupied and partially
  occupied bands, these Wannier matrices are small, which makes the
  interpolation itself very efficient. The method has been used to
  calculate the orbital magnetization of bcc Fe, hcp Co, and fcc Ni.
  Unlike an approximate calculation based on integrating orbital
  currents inside atomic spheres, our results nicely reproduce the
  experimentally measured ordering of the orbital magnetization in these
  three materials.
\end{abstract}


\maketitle

\section{Introduction}
\label{sec:intro}

Magnetism in matter originates from two distinct sources, namely, the
spin and the orbital degrees of freedom of the electrons. In many bulk
ferromagnets the spin contribution dominates, and it is therefore not
surprising that the description of spin magnetism using first-principles
methods is considerably more developed than that of orbital magnetism.
In particular, the local spin-density approximation has been successful
in studying magnetic materials for decades.\cite{Martin_04}

Although the orbital moments in bulk solids are strongly quenched by the
crystal field and typically give small contributions to the net
magnetization---between 5\% and 10\% in Fe, Co, and Ni---they can be
measured very accurately with the help of gyromagnetic
experiments.\cite{Meyer61,scott-rmp62} Moreover, there are known
instances where orbital magnetism plays a prominent role. Some examples
include weak ferromagnets with large but opposing spin and orbital
moments,\cite{Qiao_04, Gotsis_03, Taylor_02} low-coordination systems
such as magnetic nanowires,\cite{gambardella-nature02} and the
recently-predicted large orbital magnetoelectric coupling in topological
insulators and related materials.\cite{qi-prb08,essin-prl09,coh-prb11}
In addition, magnetic resonance parameters such as the
NMR\cite{Ceresoli_10a, Thonhauser_09a, Thonhauser_09b} and
EPR\cite{Ceresoli_10b} shielding tensors can be conveniently calculated
as the change in (orbital) magnetization under appropriate
perturbations.  These examples highlight the need to develop accurate
and efficient first-principles schemes for describing orbital magnetism
in solids.

The traditional way of computing the orbital magnetization $\M$ is by
integrating currents inside atom-centered muffin-tin
spheres.\cite{Wu_01,Sharma_07} This requires choosing, somewhat
arbitrarily, a cutoff radius and neglects contributions from the
interstitial regions.  A rigorous theory for the orbital magnetization
of periodic crystals free from such uncontrolled approximations was
obtained only recently\cite{Thonhauser05, Xiao05, Ceresoli06, Shi_07}
(see Ref.~\onlinecite{Thonhauser_11} for a review).  The
theory was developed in the independent-particle framework, and its
central result is the expression
\beq
\label{eq:M_def}
\begin{split}
  \M &= \frac{e}{2\hbar}\;\Im\;\sum_n\int [dk]\;f_{n\k}\\
  & \cdot\braket{\dk\unk}{\times\big(\Hk+\varepsilon_{n\k}-2\fermi\big)}
  {\dk\unk}\;.
\end{split}
\eeq
Here the integral is over the Brillouin zone (BZ), $[dk]$ stands for
$d\k/(2\pi)^3$, $\Hk=e^{-i\k\cdot\r}\ham e^{i\k\cdot\r}$ is the Bloch
Hamiltonian whose eigenfunctions are the cell-periodic Bloch functions
$\ket{\unk}$ with eigenvalues $\enk$, $f_{n\k}$ is the zero-temperature
Fermi occupation factor, and $\fermi$ is the Fermi energy.  The third
term in \equ{M_def} vanishes in ordinary insulators, but must be
included in the case of metals.\cite{Xiao05,Ceresoli06,Shi_07}

The implementation of \equ{M_def} requires a knowledge of the $\k$-space
gradients $\ket{\partial_\k \unk}$ of the occupied Bloch
states.\cite{Thonhauser_11} An easier quantity to compute in practice is
the {\it covariant derivative} $\ket{\widetilde{\partial}_\k \unk}$,
defined as the projection of $\ket{\partial_\k \unk}$ onto the
unoccupied bands.
It turns out that the replacement $\ket{\partial_\k \unk}\rightarrow
\ket{\widetilde{\partial}_\k \unk}$ in \equ{M_def} leaves the sum of its
terms invariant.  For band insulators the covariant derivative can be
conveniently evaluated by finite differences.\cite{Ceresoli06}
Unfortunately, the discretized covariant derivative approach cannot be
applied to metals, as it relies on having a constant number of occupied
bands throughout the BZ.

Thus far, the only first-principles application of \equ{M_def} to metals
is the calculation in Ref.~\onlinecite{Ceresoli_10b} of the spontaneous
orbital magnetization in Fe, Co, and Ni crystals, where the
$\k$-derivatives of the Bloch wave functions were evaluated using a
linear-response method.\cite{k.p} This carries a cost per $\k$-point
comparable to that of a non-self-consistent ground-state calculation.
The number of $\k$-points needed to converge the BZ integral in
\equ{M_def} for Fe, Co, and Ni is quite significant, rendering the full
calculation rather time-consuming.

In this work we develop an alternative approach which greatly reduces
the computational cost of evaluating \equ{M_def} for metals.  Our
implementation relies on a method for constructing well-localized
crystalline Wannier functions (WFs) by post-processing a conventional
band structure calculation.\cite{Marzari97} A key ingredient is the
``band disentanglement'' procedure of Ref.~\onlinecite{Souza01}, which
allows one to obtain a set of WFs spanning a space that contains the
occupied valence bands as a subspace.  These WFs are essentially an
exact tight-binding basis for those {\it ab initio} bands that carry
the information about $\M$.  Working in the Wannier representation, the
problem of evaluating $\M$ can then be reformulated in a very economical
way.  This reformulation involves setting up the matrix elements of
certain operators in the Wannier basis. Once that is done, the integrand
of \equ{M_def} can be evaluated very inexpensively and accurately at
arbitrary points in the BZ.  The cost per $\k$-point of the entire
procedure is significantly reduced, especially in cases where a dense
sampling of the BZ is needed to achieve convergence. Our method builds
on the work of Ref.~\onlinecite{Wang06}, where a similar ``Wannier
interpolation'' strategy was introduced to calculate the anomalous Hall
conductivity (AHC) of ferromagnetic metals.

This manuscript is organized as follows.  In Sec.~\ref{sec:prelim} the
orbital magnetization formula, \equ{M_def}, is recast in a
gauge-invariant form, and a related expression for the AHC is
introduced. We then describe step by step the formalism used to express
physical quantities in the Wannier representation.  That formalism is
applied in Sec.~\ref{sec:wr} first to the AHC, and then to the orbital
magnetization. In Sec.~\ref{sec:interpol} we describe the procedure for
evaluating the required $\k$-space matrices by Fourier interpolation.
Some details of the first-principles calculation and Wannier-function
construction are given in Sec.~\ref{sec:computational}, followed by an
application of the method to bcc Fe, hcp Co, and fcc Ni in
Sec.~\ref{sec:results}.  We conclude in Sec.~\ref{sec:conclusions} with
a brief summary and discussion.

\section{Preliminaries}
\label{sec:prelim}

\subsection{Orbital magnetization and anomalous Hall conductivity}
\label{sec:review_M_orbit}

For our purposes it will be convenient to recast \equ{M_def} in a
different form as introduced in Ref.~\onlinecite{Ceresoli06}.  We begin
by writing the axial vector $\M$ as an antisymmetric tensor,
\beq
\label{eq:M-pseudo}
\Mab=\frac{1}{2}\epsilon_{\alpha\beta\gamma}M_\gamma\;,
\eeq
where Greek indices denote Cartesian directions.  We now partition
$\Mab$ into two terms,
\beq
\label{eq:M-decomp}
\Mab = \wt{M}_{\alpha \beta}^{\rm LC}+\wt{M}_{\alpha \beta}^{\rm IC}\;.
\eeq
The ``local circulation'' is
\beq
\label{eq:LC-def}
\wt{M}_{\alpha \beta}^{\rm LC}=-\frac{e}{2\hbar}\int [dk]\,
\left[
  -2\,\Im(\Gab-\fermi\Fab)
\right],
\eeq 
and the ``itinerant circulation'' is
\beq
\label{eq:IC-def}
\wt{M}_{\alpha \beta}^{\rm IC}=-\frac{e}{2\hbar}\int [dk]\,
\left[
  -2\,\Im(\Hab-\fermi\Fab)
\right].
\eeq 
The $\k$-dependent quantities $\Fab$, $\Gab$, and $\Hab$ are 
\begin{eqnarray}
\label{eq:f-def}
\Fab&=&
\Tr\left[ (\partial_\alpha \P)\Q(\partial_\beta \P)\right]\;,\\
\label{eq:g-def}
\Gab&=&
\Tr\left[ (\partial_\alpha \P)\Q\H\Q(\partial_\beta \P)\right]\;,\\
\label{eq:h-def}
\Hab&=&
\Tr\left[ \H(\partial_\alpha \P)\Q(\partial_\beta \P)\right]\;,
\end{eqnarray}
where ``Tr'' denotes the electronic trace, $\partial_\alpha$ stands
for $\partial/\partial k_\alpha$, and the subscript $\k$ is implied in
the Bloch Hamiltonian $\H$ and in the projection operators $\P$ and
$\Q=\One-\P$ spanning the occupied and unoccupied spaces respectively
(here $\One$ is the identity operator in the full Hilbert space).

We work at $T=0$, so that Eqs.~(\ref{eq:M-pseudo})--(\ref{eq:h-def})
yield the same result as \equ{M_def}.  This can be seen by writing $\P$
in terms of the Bloch eigenstates,
\beq
\label{eq:proj-occ}
\P=\sum_n\,\ket{u_n}f_n\bra{u_n}
\eeq
and setting the occupancies $f_n$ to either one or zero.

Compared to \equ{M_def}, the above formulation has the advantage of
being manifestly {\it gauge-invariant}, i.e., independent of any
$\k$-dependent phase twists applied to the occupied Bloch states, or
more generally, any $\k$-dependent unitary mixing among them. Because
Eqs.~(\ref{eq:f-def})--(\ref{eq:h-def}) are written as traces involving
projection operators, they remain valid no matter how we choose to
represent the occupied space at each $\k$. Instead, \equ{M_def} is
written explicitly in terms of the energy eigenstates, that is, it
assumes a {\it Hamiltonian gauge}.

It should come as no surprise that it is possible to cast a physical
observable such as the orbital magnetization in a gauge-invariant form.
More interestingly, the two terms in \equ{M-decomp} are {\it
individually} gauge-invariant, and this led to the speculation that they
might be separately observable.\cite{Ceresoli06} That is indeed the
case---at least in principle---as discussed in
Ref.~\onlinecite{Souza08}.

Before continuing we mention another physical observable, the intrinsic
anomalous Hall conductivity, which can be expressed in gauge-invariant
form as
\beq
\label{eq:ahc-def}
\sigma_{\alpha\beta}^{\rm AH}=-\frac{e^2}{\hbar}
\int [dk]\,\left( -2\,\Im\,\Fab\right)\;.
\eeq
In the Hamiltonian gauge the integrand of this equation acquires a more
familiar form, i.e., as the $\k$-space Berry curvature summed over the
occupied bands.\cite{Ceresoli06}

While developing the formalism in Sec.~\ref{sec:wr} we shall first
consider the AHC before tackling the more complex case of the orbital
magnetization. This will allow us to make contact with the work of
Ref.~\onlinecite{Wang06}, where a Wannier interpolation scheme was
developed for the AHC, but using a somewhat different formulation.

\subsection{Wannier space and gauge freedom}
\label{sec:wannier-space}

The crux of our approach is to express the observables of interest
(orbital magnetization and AHC) not in terms of the Bloch eigenstates
$\ket{\unk}$, but using an alternative set of Bloch-like states
$\ket{u^\ww_{n\k}}$ constructed at each $\k$ as appropriately chosen
linear combinations of energy eigenstates. The defining feature of the
new states is that they are smooth functions of $\k$.  As a result, the
corresponding Wannier functions (WFs)
\beq
\label{eq:wf}
\ket{\R n}=\frac{1}{{\cal
    N}^3}\sum_\k\,e^{-i\k\cdot(\R-\r)}\ket{u^\ww_{n\k}} \eeq
(here $\Nq^3$ is the number of $\k$-points distributed on a uniform BZ
mesh) are well-localized in real space, and for this reason we shall say
that the states $\ket{u^\ww_{n\k}}$ belong to the {\it Wannier gauge}.
The ability to construct a short-ranged representation of the electronic
structure in real space is what will allow us to devise an efficient and
accurate interpolation scheme for the $\k$-space quantities $\Im\,\Fab$,
$\Im\,\Gab$, and $\Im\,\Hab$.

The construction of a Wannier basis proceeds in two steps, which we call
``space selection'' and ``gauge selection.''  In the case of insulators,
the space selection is typically obvious; we want the WFs to span just
the occupied subspace.  In $\k$-space, we represent this subspace by a
$\k$-dependent ``band projection operator'' $\P_\k$ defined as in
\equ{proj-occ} (with $f_n=1$ and 0 for occupied and empty bands
respectively).  Denoting the Wannier-space projection operator by
$\PP_\k$, we can then set $\PP_\k=\P_\k$.

For metals, on the other hand, the space selection step, also known as
``band disentanglement,'' is more subtle.  One wants to choose a
$\nwann$-dimensional manifold, represented by the projection operator
$\PP_\k$, throughout the BZ such that it has the following properties:
(i) it must contain as a subspace the set of all occupied eigenstates
(hence $\nwann$ cannot be less than the highest number of occupied bands
at any $\k$); (ii) it must display a smooth variation with $\k$, in the
sense that $\PP_\k$ is a differentiable function of $\k$.  A procedure
for extracting an optimally-smooth space from a larger set of band
states was developed in Ref.~\onlinecite{Souza01}.  The resulting space
typically contains some admixture of low-lying empty states in addition
to the occupied states.

The gauge selection consists of representing the smoothly-varying space
$\PP_\k$ using a set of $\nwann$ Bloch-like states which are themselves
smooth functions of $\k$,
\beq
\label{eq:proj-space}
\PP_\k=\sum_n^J\,\ket{u^\ww_{n\k}}\bra{u^\ww_{n\k}}\;.
\eeq
From these $\ket{u^\ww_{n\k}}$ the WFs are constructed via \equ{wf}.  A
procedure for selecting an optimally-smooth Wannier gauge was developed
in Ref.~\onlinecite{Marzari97}, such that the resulting WFs are
maximally-localized in the sense of having the smallest possible
quadratic spread. The localization procedure was originally devised with
an isolated group of bands in mind (e.g., the valence bands of an
insulator), but it can be applied to any smoothly-varying Bloch manifold
of fixed dimension $\nwann$.

In the Wannier gauge the projected Hamiltonian $\myhat{\HH}_\k=\PP_\k
\H_\k \PP_\k$ takes the form of a non-diagonal $\nwann\times\nwann$
matrix,
\beq
\label{eq:proj-H}
\HH^\ww_{nm}(\k)=\braket{u^\ww_{n\k}}{\H_\k}{u^\ww_{m\k}}\;.
\eeq
We define the Hamiltonian gauge in the projected space as the gauge in
which this matrix becomes diagonal,
\beq
\label{eq:H-H}
\HH_{nm}^\hh(\k)=
U^\dagger(\k)\HH^\ww(\k)U(\k)=\overline{\varepsilon}_{n\k}\delta_{nm}\;.
\eeq
Because of the nature of the space selection step, the ``projected
eigenvalues'' $\overline{\varepsilon}_{n\k}$
agree with the true {\it ab initio} eigenvalues $\enk$ for all occupied
states, but they may differ for unoccupied states.

The unitary matrices $U(\k)$ that diagonalize $\HH^\ww(\k)$ can be
used to transform other objects between the Wannier and Hamiltonian
gauges. For example, the Bloch states transform as
\beq
\label{eq:u-H}
\ket{u^\hh_{n\k}}=\sum_m^J\,\ket{u^\ww_{m\k}}U_{mn}(\k)\;.
\eeq
The gauge-invariance of the projection operators $\PP_\k$ can be checked
explicitly by inserting the right-hand-side of \equ{u-H} in place of
$\ket{u^\ww_{n\k}}$ in \equ{proj-space}.

\subsection{Projection operators and occupation matrices}
\label{sec:proj}

\begin{figure}
\includegraphics[width=6cm]{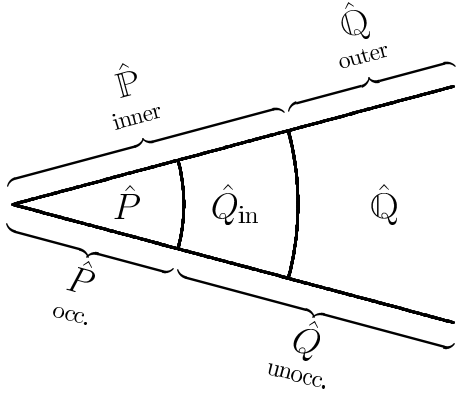}
\caption{\label{fig:spaces}The Hilbert space of Bloch functions at
wavevector $\k$ can be decomposed either as $\One=\P+\Q$ (occupied 
and unoccupied spaces) or as $\One=\PP+\QQ$ (``inner'' and ``outer'' 
spaces, where the ``inner'' space is spanned by Wannier functions).}
\end{figure}

In \equ{proj-occ} we introduced the projection operator $\P_\k$ onto
the occupied manifold at $\k$, and in \equ{proj-space} the projection
operator $\PP_\k$ onto the Wannier space at $\k$.
Figure~\ref{fig:spaces} represents schematically the relationship
between those two subspaces, as well as other related subspaces to be
defined shortly. The notation in the figure is as follows: a double
staff is used for objects that concern the distinction between the
space spanned by the WFs (``inner'') and the corresponding orthogonal
space (``outer''), while a single staff will be used for objects that
distinguish between the occupied and unoccupied parts of the Wannier
space.

Consider, for example, the projection operator $\P$ onto the occupied
bands, as defined in \equ{proj-occ}.  Recall that $\k$ labels were
suppressed in Sec~\ref{sec:review_M_orbit}; with $\k$ temporarily
restored, we have
\beq
\begin{split}
  \P_\k &= \sum_n^\nwann \ket{u^\hh_{n\k}} \,f^\hh_{n\k}\, \bra{u^\hh_{n\k}} \\
  &= \sum_{mn}^\nwann \ket{u^\ww_{m\k}} \,f^\ww_{mn,\k}\,
  \bra{u^\ww_{n\k}}\;,
\end{split}
\eeq
where $f^\ww_{mn,\k}=\bra{u^\ww_{m\k}}\P_\k\ket{u^\ww_{n\k}}$ is the
(non-diagonal) occupation matrix in the Wannier gauge.  In the following,
we will use a strongly condensed notation, leaving band indices (and
sums over them) implicit, omitting wavevector subscripts, and dropping
the superscript ``W'' from objects in the Wannier gauge.  So, for
example, we will write the Wannier-gauge Bloch states simply as
$\ket{u}$ instead of $\ket{u^\ww_{n\k}}$.

In this notation, $\P_\k$ is expressed in the Wannier gauge as just
\beq
\label{eq:P-W}
\P=\ket{u}f\bra{u}\;.
\eeq
Similarly, the projector onto the Wannier (``inner'') space is
\beq
\PP=\ket{u}\bra{u}\;.
\eeq
We also define
\begin{eqnarray}
\QQ &=& \One-\PP\;,\\
\label{eq:Q-def}
\Q &=& \One-\P=\Q_{\rm in}+\QQ\;,
\end{eqnarray}
with
\begin{eqnarray}
\Q_{\rm in} &=& \ket{u}g\bra{u}\;,\\
g &=& 1-f\;,
\end{eqnarray}
where ``1'' denotes the $\nwann\times\nwann$ identity matrix.

In practice the occupation matrix is first evaluated in the Hamiltonian
gauge in which it is diagonal, and then rotated to the Wannier gauge
with the help of \equ{u-H},
\beq
\label{eq:f}
f=Uf^\hh U^\dagger\;.
\eeq

The matrices $f$ and $g$ are idempotent, and satisfy
\beq
\label{eq:fg}
fg=gf=0\;,
\eeq
as well as
\beq
\label{eq:fH-commutator}
[f,\HH]=[g,\HH]=0\;,
\eeq
where
\beq
\label{eq:HH}
\HH=\braket{u}{\H}{u}   
\eeq
is \equ{proj-H} using the concise notation.  Equations~(\ref{eq:fg}) and
(\ref{eq:fH-commutator}) imply
\beq
\label{eq:fHg}
f\HH g=0\;.
\eeq
Note that $f\HH=\HH f=f\HH f$.  We shall also make frequent use of
relations such as $\Q_{\rm in}\QQ=\QQ Q_{\rm in}=0$, $\QQ^2=\QQ$,
$\Q_{\rm in}^2=\Q_{\rm in}$, etc.

\subsection{Compendium of ``geometric'' matrix objects}
\label{sec:compendium}

We list here for future reference a number of additional
$\nwann\times\nwann$ matrices that will be used to express $\Im\,\Fab$,
$\Im\,\Gab$, and $\Im\,\Hab$ in the Wannier representation:
\begin{eqnarray}
\label{eq:AA}
\AA_\alpha&=&i\me{u}{\partial_\alpha u}\;,\\
\label{eq:FF}
\FF_{\alpha\beta}&=&\me{\partial_\alpha u}{\partial_\beta u}\;,\\
\label{eq:FFtil}
\wt{\FF}_{\alpha\beta}&=&\braket{\partial_\alpha u}{\QQ}{\partial_\beta u}
=\FF_{\alpha\beta}-\AA_\alpha\AA_\beta\;,\\
\label{eq:OO}
\OO_{\alpha\beta}&=&i\FF_{\alpha\beta}-i\FF_{\alpha\beta}^\dagger
                 =\partial_\alpha\AA_\beta - \partial_\beta\AA_\alpha\;,\\
\label{eq:OOtil}
\wt{\OO}_{\alpha\beta}&=&
i\wt{\FF}_{\alpha\beta}-i\wt{\FF}_{\alpha\beta}^\dagger
=\OO_{\alpha\beta}-i\left[ \AA_\alpha,\AA_\beta \right]\;.
\end{eqnarray}
The Hermitian matrices $\AA_\alpha$, $\OO_{\alpha\beta}$, and
$\wt{\OO}_{\alpha\beta}$ are known as the Berry connection, Berry
curvature, and gauge-covariant Berry curvature.  They play a central
role in the theory of geometric-phase effects in solids.\cite{Xiao10}

In addition to the above objects, which represent intrinsic properties
of the Bloch manifold, we shall make use of a number of
similarly-defined quantities which also depend on the Hamiltonian (and
therefore are not strictly speaking ``geometric''):
\begin{eqnarray}
\label{eq:BB}
\BB_\alpha&=&i\braket{u}{\H}{\partial_\alpha u}\;,\\
\label{eq:BBtil}
\wt{\BB}_\alpha&=&i\braket{u}{\H\QQ}{\partial_\alpha u}
=\BB_\alpha-\HH\AA_\alpha\;,\\
\label{eq:CC}
\CC_{\alpha\beta}&=&\braket{\partial_\alpha u}{\H}{\partial_\beta u}\;,\\
\label{eq:CCtil}
\wt{\CC}_{\alpha\beta}&=&
\braket{\partial_\alpha u}{\QQ \H\QQ}{\partial_\beta u}\nn
&=&\CC_{\alpha\beta}-\AA_\alpha\wt{\BB}_\beta-\BB^\dagger_\alpha\AA_\beta\;,\\
\label{eq:LL}
\LL_{\alpha\beta}&=&i\CC_{\alpha\beta}-i\CC^\dagger_{\alpha\beta}\;,\\
\label{eq:LLtil}
\wt{\LL}_{\alpha\beta}&=&
i\wt{\CC}_{\alpha\beta}-i\wt{\CC}^\dagger_{\alpha\beta}\nonumber\\
&=&\LL_{\alpha\beta}
-i\left( \AA_\alpha\wt{\BB}_\beta-\AA_\beta \BB_\alpha - \text{h.c.} \right),
\end{eqnarray}
where ``h.c.'' stands for Hermitian conjugate. Note that while
$\LL_{\alpha\beta}$ and $\wt{\LL}_{\alpha\beta}$ are Hermitian,
$\BB_\alpha$ is not.

All of the matrices listed above are written in the Wannier gauge, where
they are smooth functions of $\k$ and can be evaluated efficiently using
Fourier transforms, as will be described in Sec.~\ref{sec:interpol}.

\section{Wannier-space representation of physical quantities}
\label{sec:wr}

\subsection{Anomalous Hall conductivity}
\label{sec:ahc}

\subsubsection{Derivation}

As a first application of our formalism, let us consider \equ{ahc-def}
for the AHC. The integrand is the Berry curvature $-2\,\Im\,\Fab$, and we
wish to write it as a trace of products of matrices defined within the
Wannier space.

Our starting point is \equ{f-def} for $\Fab$. In preparation for
taking the trace therein, let us express $(\partial_\alpha\P)\Q$ in
the Wannier gauge. Differentiating \equ{P-W} leads to
\beq
\partial_\alpha\P=\ket{\partial_\alpha u}f\bra{u}+
\ket{u}f\bra{\partial_\alpha u}+
\ket{u}f_\alpha\bra{u}\;,
\eeq
where $f_\alpha=\partial_\alpha f$.  Multiplying on the right with
$\Q$ and using \equ{Q-def} yields two terms for
$(\partial_\alpha\P)\Q$. One is
\beq
\label{eq:dPQ1}
(\partial_\alpha \P)\QQ=\ket{u}f\bra{\partial_\alpha u}\QQ\;,
\eeq
where $\bra{u}\QQ=0$ was used, and the other is
\beq
\label{eq:dPQ2}
(\partial_\alpha \P)\Q_{\rm in}=\ket{u}f(+i\AA_\alpha)g\bra{u}+
\ket{u}f_\alpha g\bra{u}\;,
\eeq
where \equ{fg} was used. Now, from \equ{f} we find\cite{footnote1}
\beq
\label{eq:f_a}
f_\alpha=i\left[f, J_\alpha\right]\;,
\eeq
where the Hermitian matrix $J_\alpha$, like $f$ itself, is first
evaluated in the Hamiltonian gauge, being defined as
\beq 
\label{eq:J-H}
J_\alpha^\hh=iU^\dagger \partial_\alpha U\;,
\eeq
and then rotated into the Wannier gauge,
\beq
\label{eq:J}
J_\alpha = UJ_\alpha^\hh U^\dagger\;.
\eeq

Using \equ{f_a} in \equ{dPQ2} and defining
\beq
\label{eq:a}
A_\alpha=\AA_\alpha+J_\alpha\;,
\eeq
we arrive at the compact form
\beq
\label{eq:dPQ2b}
(\partial_\alpha \P)\Q_{\rm in}=i\ket{u}fA_\alpha g\bra{u}\;.
\eeq
The desired expression is the sum of Eqs.~(\ref{eq:dPQ1}) and
(\ref{eq:dPQ2b}),
\beq
\label{eq:dPQ}
(\partial_\alpha\P)\Q=\ket{u}f\bra{\partial_\alpha u}\QQ
+i\ket{u}fA_\alpha g\bra{u}\;.
\eeq

With this relation in hand it becomes straightforward to evaluate
\equ{f-def}, and we quickly arrive at
\beq
\Fab=\tr\left[ f\wt{\FF}_{\alpha\beta}
+f A_\alpha g A_\beta \right],
\eeq
where ``tr'' denotes the trace over $\nwann\times \nwann$ matrices, not
to be confused with the trace ``Tr'' over the full Hilbert space
introduced earlier.  We are interested in the imaginary part, and using
\equ{OOtil} we find
\begin{eqnarray}
\label{eq:imf-a}
-2\,\Im\,\Fab&=&\Re\,\tr[ f\wt{\OO}_{\alpha\beta}]
-2\,\Im\,\tr[fA_\alpha g A_\beta]\\
&=&\Re\,\tr[ f\OO_{\alpha\beta}]
+2\,\Im\,\tr[f\AA_\alpha\AA_\beta-fA_\alpha g A_\beta]\;.\nonumber
\end{eqnarray}
Expanding $A_\alpha$ and using the identity
\begin{eqnarray}
\Im\,\tr[f\AA_\alpha\AA_\beta]&=& \Im\,\tr[f\AA_\alpha(f+g)\AA_\beta]\\\nonumber
&=& \Im\,\tr[f\AA_\alpha g\AA_\beta]\;,
\end{eqnarray}
we end up with
\begin{eqnarray}
\label{eq:imf-b}
-2\,\Im\,\Fab&=&\Re\,\tr\left[ f\OO_{\alpha\beta}\right]\\
&-&2\,\Im\,\tr[ f\AA_\alpha g J_\beta +fJ_\alpha g\AA_\beta+
fJ_\alpha gJ_\beta ]\;.\nonumber
\end{eqnarray}
This expression for the Berry curvature of the occupied states is our
first important result.

\subsubsection{Discussion}
\label{sec:ahc-disc}

The ingredients that go into Eq.~(\ref{eq:imf-b}) are the matrices
$\AA_\alpha$, $\OO_{\alpha\beta}$, $f$, and $J_\alpha$ expressed in a
smooth (but otherwise arbitrary) gauge. It should be noted that while
$\AA_\alpha$ and $\OO_{\alpha\beta}$ are themselves smooth functions of
$\k$, this is not so for $f$ and $J_\alpha$. Consider $f$, given by
\equ{f}. In metals it is affected by the step-like discontinuity in
$f^\hh$ at the Fermi surface. More generally it is also affected by the
wrinkles in the rotation matrix $U$ (recall that $U$ relates via
\equ{u-H} the smooth Wannier-gauge Bloch states to the Hamiltonian-gauge
states, which are non-analytic as a function of $\k$ at points of
degeneracy). The situation is even more severe in the case of
$J_\alpha$. Because it contains the derivative $\partial_\alpha U$ of a
non-smooth function, it has spikes in $\k$-space.

How does one reconcile the existence of irregular and spiky quantities
inside \equ{imf-b} with the form of \equ{f-def}, which suggests that
$\Fab$ is a smooth function of $\k$, except possibly when crossing the
Fermi surface (when a state comes in or out of the occupied manifold,
introducing a discontinuity in $\P_\k$)?  The answer is that while
$J_\alpha$ itself has a very irregular behavior, combinations like
$fJ_\alpha g$ which actually appear in \equ{imf-b} do not, as will be
discussed in Sec.~\ref{sec:fourier}.

Let us now make contact with the formulation of Wannier interpolation
developed in Ref.~\onlinecite{Wang06}.  In that work the expression
for the Berry curvature of the occupied states [Eq.~(32) therein] was
written in the Hamiltonian gauge, where the occupation matrices are
diagonal. This required transforming $\AA_\alpha$ and
$\OO_{\alpha\beta}$ from the Wannier gauge, where they are first
constructed, to the Hamiltonian gauge. Here we choose to keep
everything in the Wannier gauge throughout.  The advantage is that
even though the matrices $f$ and $J_\alpha$ have to be constructed
first in the Hamiltonian gauge, it is straightforward to rotate them
into the Wannier gauge where all other needed objects are constructed.
Instead, the reverse transformation of those other objects can in
certain cases become nontrivial.

The two formulations are of course equivalent, and it is instructive to
recover explicitly from \equ{imf-b} the corresponding expression in
Ref.~\onlinecite{Wang06}. Consider for example the last term in
\equ{imf-b}. Taking the trace in the Hamiltonian gauge we find
\beq
\label{eq:recover1}
\tr [f J_\alpha g J_\beta]=
\sum_{nm}^\nwann\, f_n^\hh J^\hh_{\alpha,nm}(1-f_m^\hh)J^\hh_{\beta,mn}\;,
\eeq
and thus
\beq
\label{eq:recover3}
-2\,\Im\,\tr [f J_\alpha g J_\beta]=
-i\sum_{nm}^\nwann\,(f_m^\hh-f_n^\hh)J^\hh_{\alpha,nm}J^\hh_{\beta,mn}\;,
\eeq
which agrees with the last term in Eq.~(32) of Ref.~\onlinecite{Wang06}
($D_\alpha^{(\hh)}$ therein corresponds in our notation to
$-iJ^\hh_\alpha$).

It is pleasing to see that \equ{imf-b}, when converted to the
Hamiltonian gauge, reduces to what was termed in
Ref.~\onlinecite{Wang06} the ``sum over occupied bands'' expression,
where individual terms have spiky features only when two bands, one
occupied the other empty, almost touch at the Fermi level, as the factor
$f_n^\hh-f_m^\hh$ suppresses spikes associated with pairs of occupied
states. As we shall see in Sec.~\ref{sec:fourier}, this is a general
feature of our formulation, which leads naturally to expressions where
the spiky object $J_\alpha$ appears under the trace sandwiched between
$f$ and $g$.

\subsection{Orbital magnetization}
\label{sec:orb}

Let us now apply to $\Im\,\Hab$ and $\Im\,\Gab$ the same strategy
developed above for $\Im\,\Fab$, completing the list of quantities
needed to evaluate the orbital magnetization.  We remark that it would
be possible to proceed along the lines of Ref.~\onlinecite{Wang06} in
order to arrive at ``sum over occupied bands'' expressions for those
quantities in the Hamiltonian gauge. However, we found such an approach
to be rather cumbersome, especially in the case of $\Im\,\Gab$.

\subsubsection{Derivation}
\label{sec:orb-der}

Inserting \equ{dPQ} into \equ{h-def} leads to
\beq
\Hab=\tr \left[f\HH f\wt{\FF}_{\alpha\beta} +f\HH f A_\alpha g A_\beta\right],
\eeq
so that
\begin{eqnarray}
-2\,\Im\,\Hab&=&\Re\,\tr \left[f\HH f\wt{\OO}_{\alpha\beta}\right]
-2\,\Im\,\tr\left[ f\HH f A_\alpha g A_\beta \right]\nn
&=&\Re\,\tr \left[f\HH f\OO_{\alpha\beta}\right]\nn
&+&2\,\Im\,\tr \left[ f\HH f\AA_\alpha\AA_\beta - f\HH f A_\alpha g A_\beta \right].
\end{eqnarray}
Using \equ{a}, this takes the desired form
\beq
\label{eq:imh-tr}
\begin{split}
-2\,\Im\,\Hab&=\Re\,\tr\left[ f\HH f \OO_{\alpha\beta}\right]
+2\,\Im\,\tr\left[ f\HH f \AA_\alpha f\AA_\beta\right]\\
&-2\,\Im\,\tr
\left[ 
  f\HH f
  \left( \AA_\alpha g J_\beta +J_\alpha g\AA_\beta+J_\alpha gJ_\beta\right)
\right]
\end{split}
\eeq
in terms of basic matrix objects with every $J_\alpha$
sandwiched between an $f$ and a $g$ (after taking account of the
cyclic property of the trace).

Repeating for $\Gab$, we find
\beq
\Gab=\tr
\left[
  f\left(
    \wt{\CC}_{\alpha\beta}+
    A_\alpha g\wt{\BB}_\beta
    +\wt{\BB}^\dagger_\alpha gA_\beta+A_\alpha g\HH g A_\beta
    \right)
\right]
\eeq
and
\begin{eqnarray}
\lefteqn{-2\,\Im\,\Gab=\Re\,\tr\left[
f\wt{\LL}_{\alpha\beta}\right]}\\
&&{}-2\,\Im\,\tr
\left[
  \left( fA_\alpha g\wt{\BB}_\beta-\alpha\leftrightarrow\beta \right)
  +fA_\alpha g\HH g A_\beta
\right]\;.\nonumber
\end{eqnarray}
Expanding $\wt{\LL}_{\alpha\beta}$ and $\wt{\BB}_\beta$, this becomes
\begin{eqnarray}
\lefteqn{-2\,\Im\,\Gab=\Re\,\tr\left[
f\LL_{\alpha\beta}\right]}\nn
&&{}-2\,\Im\,\tr
\Big[
  f\left(
    A_\alpha g\BB_\beta-A_\alpha g\HH\AA_\beta-\AA_\alpha\BB_\beta
    -\alpha\leftrightarrow\beta
    \right)\nn
    &&{}+f\left(
      \AA_\alpha\HH\AA_\beta+A_\alpha g\HH g A_\beta
      \right)
\Big]\;.
\end{eqnarray}

Next we expand $A_\alpha$ and gather terms in three groups, containing
zero, one, and two occurrences of the matrices $J_\alpha$ and
$J_\beta$,\cite{footnote2}
\beq
\label{eq:J0J1J2}
-2\,\Im\,\Gab=J0+J1+J2\;.
\eeq
The $J2$ group contains only one term,
\beq
J2=-2\,\Im\,\tr\left[ fJ_\alpha g\HH g J_\beta\right]\;.
\eeq
The $J1$ group is
\begin{eqnarray}
J1&=&-2\,\Im\,\tr
\left[
  fJ_\alpha g
  \left(
    \BB_\beta-\HH\AA_\beta+\HH g\AA_\beta
  \right)
  -\alpha\leftrightarrow\beta
\right]\nn
&=&-2\,\Im\,\tr
\left[
  fJ_\alpha g\BB_\beta-\alpha\leftrightarrow\beta
\right]\;,
\end{eqnarray}
where in the second equality we replaced one instance of $g$ with $1-f$
and then used \equ{fHg}.  The $J0$ group reads, after combining certain
terms and canceling out some others,
\begin{eqnarray}
\lefteqn{J0=\Re\,\tr\left[ f\LL_{\alpha\beta}\right]}\\
&&{}+2\,\Im\,\tr
\left[
  f\left(
    \AA_\alpha f\BB_\beta-\AA_\beta f\BB_\alpha-\AA_\alpha\HH f\AA_\beta
    \right)
\right]\;.\nonumber
\end{eqnarray}
This can be simplified further with the help of the following identity
proven in the Appendix,
\beq
\label{eq:identity}
\tr\left[ f\AA_\alpha f\BB_\beta\right]=
\tr\left[ f\AA_\alpha f\HH\AA_\beta\right]\;,
\eeq
which leads to
\beq
\label{eq:J0}
J0
=\Re\,\tr\left[ f\LL_{\alpha\beta}\right]
-2\,\Im\,\tr\left[ f\HH f\AA_\alpha f\AA_\beta\right].
\eeq
Collecting terms, we find
\begin{eqnarray}
\label{eq:img-tr}
-2\,\Im\,\Gab&=&\Re\,\tr\left[f\LL_{\alpha\beta}\right]
-2\,\Im\,\tr\left[ f\HH f\AA_\alpha f\AA_\beta\right]\\
&-&2\,\Im\,\tr
\left[
  f\left(
    J_\alpha g\BB_\beta-\alpha\leftrightarrow\beta
  \right)
  +fJ_\alpha g\HH gJ_\beta
\right]\;.\nonumber
\end{eqnarray}

\subsubsection{Final expressions}
\label{sec:orb-final}

The quantities $\Im\,\Fab$, $\Im\,\Gab$, and $\Im\,\Hab$ enter the
orbital magnetization expression in the combinations $\Im(\Gab-\fermi
\Fab)$ and $\Im(\Hab-\fermi \Fab)$.  Using the condensed notations
\begin{eqnarray}
X^\lo&=&fXf\;,\\
\label{eq:Xhi}
X^\hi&=&gXg\;,\\
X^+&=&gXf\;,\\
X^-&=&fXg\;,
\end{eqnarray}
in Eqs.~(\ref{eq:imf-b}), (\ref{eq:imh-tr}), and (\ref{eq:img-tr}), we
obtain for the integrand of $\wt{M}_{\alpha \beta}^{\rm IC}$,
\equ{IC-def},
\begin{eqnarray}
\label{eq:ic-final}
\lefteqn{-2\,\Im(\Hab-\fermi\Fab)=}\nn
&&{}+\Re\,\tr\left[ \left(\HH^\lo-\fermi\right)\OO^\lo_{\alpha\beta}\right]
+2\,\Im\,\tr\left[ \HH^\lo\AA^\lo_\alpha\AA^\lo_\beta\right]\\
&&{}-2\,\Im\,\tr
\left[
  \left(\HH^\lo-\fermi\right)
 \left( \AA^-_\alpha J^+_\beta+J^-_\alpha\AA^+_\beta+J^-_\alpha J^+_\beta\right)
\right]\;,\nonumber
\end{eqnarray}
and for the integrand of $\wt{M}_{\alpha \beta}^{\rm LC}$, \equ{LC-def},
\begin{eqnarray}
\label{eq:lc-final}
\lefteqn{-2\,\Im(\Gab-\fermi\Fab)=}\nn
&&{}+\Re\,\tr\left[\LL^\lo_{\alpha\beta}-\fermi\OO^\lo_{\alpha\beta}\right]
-2\,\Im\,\tr\left[ \HH^\lo\AA^\lo_\alpha\AA^\lo_\beta\right]\\
&&{}-2\,\Im\,\tr
\Big\{
  \big[ 
    J^-_\alpha\big(\BB^+_\beta-\fermi\AA^+_\beta\big)
    -\alpha\leftrightarrow\beta
  \big]\nn
&&{}+J^-_\alpha(\HH^\hi-\fermi)J^+_\beta\Big\}\;.\nonumber
\end{eqnarray}
Note that the second terms in these two equations are equal and
opposite.  For the special case of an insulator with $f=1$ and $g=0$,
only the first two terms are nonzero in each of
Eqs.~(\ref{eq:ic-final}-\ref{eq:lc-final}), and these expressions reduce
to those derived in Ref.~\onlinecite{Ceresoli06}.

\section{Interpolation of the Wannier-gauge matrices}
\label{sec:interpol}

We calculate the orbital magnetization by averaging
\eqs{ic-final}{lc-final} over a sufficiently dense grid of $\k$-points in
order to approximate the BZ integrals in \eqs{LC-def}{IC-def}. At each
$\k$ the matrices $f$, $J_\alpha$, $\HH$, $\AA_\alpha$, $\BB_\alpha$,
$\OO_{\alpha\beta}$, and $\LL_{\alpha\beta}$ are needed, and they are
calculated by Fourier interpolation as follows.

\subsection{Fourier transform expressions}
\label{sec:fourier}

We start with the matrix $\HH$. Inverting \equ{wf}, \beq
\label{eq:bloch-sum}
\ket{u}=\sum_\R\,e^{-i\k\cdot(\r-\R)}\ket{\R}\;,
\eeq
and inserting into \equ{HH} yields
\beq
\label{eq:HH-R}
\HH=\sum_\R\,\eikr\braket{\0}{\ham}{\R}\;.
\eeq
We emphasize that any desired wavevector $\k$ can be plugged into this
expression, allowing one to smoothly interpolate the matrix $\HH$ between
the $\Nq^3$ {\it ab initio} grid points used in \equ{wf} to construct
the WFs. Diagonalizing $\HH$ [\equ{H-H}] and using the resulting
rotation matrix $U$ and interpolated eigenvalues
$\overline{\varepsilon}_n$ in \equ{f} yields the occupation matrix
$f$.

Next we consider the matrix $J_\alpha$.  In practice it can be
calculated by inserting $U$ into \equ{J} and then using\cite{Wang06}
\beq
\label{eq:J-fourier}
J^\hh_{\alpha,nm}=
\begin{cases}
\displaystyle
\frac{i\left[ U^\dagger\HH_\alpha U\right]_{nm}}
{\overline{\varepsilon}_m-\overline{\varepsilon}_n} & \text{if $n\not= m$}\\
0 & \text{if $n=m$}
\end{cases},
\eeq
where $\HH_\alpha=\partial_\alpha\HH$ is obtained by differentiating
\equ{HH-R}.  In the vicinity of band degeneracies and weak avoided
crossings the denominator in \equ{J-fourier} becomes small, leading to
strong peaks in $J^\hh_{\alpha,nm}$ as a function of $\k$.  If both
bands $n$ and $m$ are occupied, such peaks must eventually cancel out in
the final expressions for the AHC and orbital magnetization, as
discussed in Sec.~\ref{sec:ahc-disc}.

We can make such cancellations explicit from the outset by noting that
$J_\alpha$ only appears in the combinations $J^-_\alpha=fJ_\alpha g$ and
$J^+_\alpha=gJ_\alpha f$.  Taking the former, for example, we find using
\eqs{f}{J} that
\beq
\label{eq:J-pm}
J^-_\alpha
=UJ_\alpha^{{\rm H}-} U^\dagger
\eeq
where the matrix elements of $J^{{\rm H}-}_\alpha=f^\hh
J^\hh_\alpha\left( 1-f^\hh\right)$ are
\beq
\label{eq:Jminus-fourier}
J^{{\rm H}-}_{\alpha,nm}=
\begin{cases}
  \displaystyle \frac{i\left[ U^\dagger\HH_\alpha U\right]_{nm}}
  {\overline{\varepsilon}_m-\overline{\varepsilon}_n} & 
\text{if $n$ occupied and $m$ empty,}\\
  0 & \text{otherwise.}
\end{cases}
\eeq
$J^{{\rm H}+}_{\alpha,nm}$ is given by a similar expression, but with
$m$ occupied and $n$ empty.  Unlike \equ{J-fourier}, the expressions for
$J^{{\rm H}-}$ and $J^{{\rm H}+}_\alpha$ are well behaved in that they
will only show peaks when the direct gap is small and mixing between
occupied and empty states is strong. By working directly with them, we
avoid introducing any quantity that would react strongly to crossings
among occupied states.

While $f$, $J_\alpha^-$ and $J_\alpha^+$ are first calculated in the
Hamiltonian gauge and then converted to the Wannier gauge, the remaining
quantities entering \eqs{ic-final}{lc-final} are most easily calculated
directly in the Wannier gauge, in the same way as $\HH$.  It is
sufficient to consider the Wannier representation of the three basic
quantities $\AA_\alpha$, $\BB_\alpha$, and $\CC_{\alpha\beta}$
introduced in Sec.~\ref{sec:compendium}.  Inserting \equ{bloch-sum} into
the respective definitions we find
\begin{eqnarray}
\label{eq:AA-R}
\AA_\alpha&=&\sum_\R\,\eikr\braket{\0}{\rop_\alpha}{\R}\;,\\
\label{eq:BB-R}
\BB_\alpha&=&\sum_\R\,\eikr\braket{\0}{\ham(\rop-R)_\alpha}{\R}\;,\\
\label{eq:CC-R}
\CC_{\alpha\beta}&=&\sum_\R\,\eikr\braket{\0}{\rop_\alpha\ham(\rop-R)_\beta}{\R}\;.
\end{eqnarray}
The expressions for $\OO_{\alpha\beta}$ and $\LL_{\alpha\beta}$ are
obtained by inserting \equ{AA-R} into \equ{OO} and \equ{CC-R} into
\equ{LL}, respectively.

It was shown in Ref.~\onlinecite{Wang06} that the computation of the AHC
requires a knowledge of the Wannier matrix elements of $\ham$ and
$\hat{\r}$.  Inspection of the Fourier transform expression given above
reveals that the bulk orbital magnetization requires in addition the
matrix elements of $\ham\hat{\r}$ and $\hat{\r}\ham\hat{\r}$.  This is
more than might have been anticipated, given that the matrix elements of
$\hat{\r}$ and $\ham\hat{\r}$ are not needed for calculating the orbital
moment of finite samples under open boundary conditions, but it is the
price to be paid for a formulation that extends also to the case of
periodic boundary conditions.

\subsection{Evaluation of the real-space matrices}
\label{sec:real-space}

We shall follow the approach of Ref.~\onlinecite{Wang06}, whereby the
needed real-space matrix elements are actually evaluated in reciprocal
space.  Inverting the Fourier sums in Eqs.~(\ref{eq:HH-R}),
(\ref{eq:AA-R}), (\ref{eq:BB-R}), and (\ref{eq:CC-R}), we find
\begin{eqnarray}
\label{eq:H}
\braket{\0}{\ham}{\R}&=&\frac{1}{\Nq^3}\sum_\k\,\emikr\HH_\k\;,\\
\label{eq:A}
\braket{\0}{\rop_\alpha}{\R}&=&\frac{i}{\Nq^3}\sum_\k\,\emikr
\me{u_\k}{\partial_\alpha u_\k}\nonumber\\
&\simeq&\frac{i}{\Nq^3}\sum_{\k,\b}\,\emikr
w_b b_\alpha\MM_{\k\;\b}\;,\\
\label{eq:B}
\braket{\0}{\ham(\rop-R)_\alpha}{\R}&=&\frac{i}{\Nq^3}\sum_\k\,\emikr 
\braket{u_\k}{\H_\k}{\partial_\alpha u_\k}\nonumber\\
&\simeq&\frac{i}{\Nq^3}\sum_{\k,\b}\,\emikr w_b b_\alpha\HH_{\k,\b}\;,\\
\label{eq:C}
\braket{\0}{\rop_\alpha\ham(\rop-R)_\beta}{\R}&=&\frac{1}{\Nq^3}\sum_\k\,\emikr
\braket{\partial_\alpha u_\k}{\H_\k}{\partial_\beta u_\k}\nonumber\\
\simeq\frac{1}{\Nq^3}\sum_{\k,\b_1,\b_2}\mspace{-35mu}&&\emikr
w_{b_1}b_{1\alpha}w_{b_2}b_{2\beta}\HH_{\k,\b_1,\b_2}\;,\nonumber\\
\end{eqnarray}
where the sums are over the $\Nq^3$ points in the uniform {\it ab
  initio} mesh. The second equalities in
Eqs.~(\ref{eq:A})--(\ref{eq:C}) follow from using a finite-difference
expression for the derivatives of the smooth Bloch
states,\cite{Marzari97}
\beq
\label{eq:disc}
\ket{\da u_\k}=\sum_\b w_b b_{\alpha}\ket{u_{\k+\b}}+{\cal O}(b^2)
\eeq
($w_b$ are appropriately chosen weights, and the sum is over shells of
vectors $\b$ connecting a point $\k$ on the {\it ab initio} grid to its
neighbors), together with the definitions
\begin{eqnarray}
\label{eq:HH-b}
\HH_{\k,\b}&=&\braket{u_\k}{\H_\k}{u_{\k+\b}}\;,\\
\label{eq:MM}
\MM_{\k,\b}&=&\me{u_\k}{u_{\k+\b}}\;,\\
\label{eq:HH-b2}
\HH_{\k,\b_1,\b_2}&=&\braket{u_{\k+\b_1}}{\H_\k}{u_{\k+\b_2}}\;,
\end{eqnarray}
which complement \equ{HH} for $\HH_\k$, needed in \equ{H}. Writing the
states $\ket{u}$ as linear combinations of the original {\it ab initio}
eigenstates $\ket{u^0}$,
\beq
\label{eq:proj}
\ket{u_{n\k}} = \sum_m^{\nbands_\k} \ket{u^0_{m\k}} V_{\k,mn}\;,
\eeq
we arrive at the following expressions for the $\nwann\times\nwann$
matrices appearing in Eqs.~(\ref{eq:H})--(\ref{eq:C}):
\begin{eqnarray}
\HH_\k&=&V_\k^\dagger{\cal H}_\k V_\k\;,\\
\HH_{\k,\b} &=& 
V_\k^\dagger {\cal H}_\k {\cal M}_{\k,\b}V_{\k+\b}\;,\\
\MM_{\k,\b}&=&V_\k^\dagger{\cal M}_\k V_{\k+\b}\;,\\
\HH_{\k,\b_1,\b_2} &=&
V_{\k+\b_1}^\dagger {\cal H}_{\k,\b_1,\b_2}V_{\k+\b_2}\;,
\end{eqnarray}
where
\begin{eqnarray}
\label{eq:1-point}
{\cal H}_\k&=&\braket{u^0_\k}{\H_\k}{u^0_\k}\;,\\
\label{eq:2-point}
{\cal M}_{\k,\b}&=&\me{u^0_\k}{u^0_{\k+\b}}\;,\\
\label{eq:3-point}
{\cal H}_{\k,\b_1,\b_2}&=&\braket{u^0_{\k+\b_1}}{\H_\k}{u^0_{\k+\b_2}}\;.
\end{eqnarray}

The diagonal eigenvalue matrix ${\cal H}_\k$ and the overlap matrix
${\cal M}_{\k,\b}$ are readily available, as they constitute the input
to the space-selection and gauge-selection steps in the wannierization
procedure. While they suffice for calculating the AHC,\cite{Wang06} as
well as $\wt{\M}^{\rm IC}$, the $\wt{\M}^{\rm LC}$ term in the orbital
magnetization requires the additional quantities ${\cal
H}_{\k,\b_1,\b_2}$.

\section{Computational Details}
\label{sec:computational}

Planewave pseudopotential calculations were carried out for the
ferromagnetic transition metals bcc Fe, hcp Co, and fcc Ni at their
experimental lattice constants (5.42, 4.73, and 6.65~bohr,
respectively).  The calculations were performed in a noncollinear spin
framework, using fully-relativistic norm-conserving
pseudopotentials\cite{Corso05} generated from similar parameters as in
Ref.~\onlinecite{Wang06}.  The energy cutoff for the expansion of the
valence wavefunctions was set at $120$~Ry for Fe and Ni and at $140$~Ry
for Co; a cutoff of 800~Ry was used for the charge density.  Exchange
and correlation effects were treated within the PBE generalized-gradient
approximation.\cite{Perdew96}

The calculation of the orbital magnetization comprises the following
sequence of steps: (i) self-consistent total-energy calculation; (ii)
non-self-consistent band structure calculation including several
conduction bands; (iii) evaluation of the matrix elements in
\eqs{2-point}{3-point}; (iv) wannierization of the selected bands; and
(v) Wannier interpolation of \eqs{ic-final}{lc-final} across a dense
$\k$-point mesh, with the value of $\fermi$ taken from step~(i).
Steps~(i) and (ii) were carried out using the \textsc{PWscf} code
from the \textsc{Quantum-Espresso} package,\cite{Giannozzi09} and in
step~(iii) we used the interface routine \textsc{pw2wannier90} from the
same package, modified to calculate \equ{3-point} in addition to
\equ{2-point}.  Step~(iv) was done using the \textsc{Wannier90} code,
\cite{Mostofi08} and for step~(v) a new set of routines was written (we
plan to incorporate these in a future release of the \textsc{Wannier90}
distribution).

The BZ integration in step~(i) was carried out on a
16$\times$16$\times$16 Monkhorst-Pack mesh,\cite{Monkhorst76} using a
Fermi smearing of 0.02~Ry.  In step~(ii), the 28 lowest band states were
calculated for bcc Fe and fcc Ni on $\Nq\times\Nq\times\Nq$ $\k$-point
grids including the $\Gamma$-point.  (For hcp Co, with two atoms per
cell, 48 states per $\k$-point were calculated.)  After testing several
grid densities for convergence (see Sec.~\ref{sec:convergence} below),
we settled on $\Nq=10$ for all three materials.  In step~(iv) we
followed the procedure described in Ref.~\onlinecite{Wang07} to generate
eighteen disentangled spinor WFs per atom, capturing the $s$, $p$, and
$d$ characters of the selected bands. In the case of fcc Ni, we also
tested an alternative set consisting of only fourteen WFs, ten of which
are atom-centered $d$-like orbitals while the remaining four are
$s$-like and are centered at the tetrahedral interstitial
sites.\cite{Souza01} We found excellent agreement --~to within
0.0002~$\mu_B$/atom~-- between the values of $\M$ obtained with the two
sets of WFs.

It should be kept in mind that our calculations use a pseudopotential
framework in which the contributions to $\M$ coming from the core region
are not quite described correctly.  A rigorous treatment using the
so-called GIPAW approach~\cite{pickard-prb01} was developed in
Refs.~\onlinecite{Ceresoli_10a,Ceresoli_10b}. It was shown that
\equ{M_def}, written in terms of the pseudo-wavefunctions and
pseudo-Hamiltonian, must be supplemented by certain core-reconstruction
corrections (CRCs) in order to obtain the full orbital magnetization.

We know from the work of Ref.~\onlinecite{Ceresoli_10b} that the CRCs
are small for bulk Fe, Co, and Ni, of the order of 5\%.  This suggests
that the errors inherent in our uncontrolled approximations in the core
are also of the same order.  If one wants to treat the problem correctly
and capture this missing 5\%, one should use the GIPAW approach.
However, the issues of implementing the CRCs are completely orthogonal
to the issues of Wannier interpolation, and so we have not pursued that
here. (As the CRCs originate in the atomic cores, there is nothing to be
gained from using Wannier interpolation to calculate them.)
Alternatively, the Wannier matrix elements could be generated starting
from an all-electron first-principles calculation,\cite{freimuth-prb08}
in which case the present formulation should yield the full
first-principles orbital magnetization.

Finally, we mention an issue in all DFT-based studies of orbital-current
effects, namely that the accuracy of the ordinary exchange-correlation
functionals (LSDA, GGA, GGA+U, etc.) has not been well tested in this
context.  A variety of interesting ideas have been proposed for improved
functionals,
\cite{wensauer-prb04,rohra-prl06,morbec-ijqc08,abedinpour-2009} but
exploring these would take us outside the scope of the present work.

\section{Results}
\label{sec:results}

In this section, the Wannier interpolation method is used to calculate
the orbital magnetization of the ferromagnetic transition metals bcc Fe,
hcp Co, and fcc Ni. We begin by carrying out convergence tests with
respect to BZ sampling. Converged values are then tabulated for the
three materials, and compared with measurements and previous
calculations. Finally, we investigate how $\M$ is distributed in
$\k$-space in the case of bcc Fe.

\subsection{Convergence studies}
\label{sec:convergence}

Recall that two separate BZ grids are employed at different stages of
the calculation (the {\it ab initio} grid used to evaluate the Wannier
matrix elements, and the interpolation grid used to carry out the BZ
integrals in the orbital magnetization expression), and both must be
checked for convergence.

Table~\ref{tab:convergence} shows the calculated orbital magnetization
as a function of the number $N\times N\times N$ of points on a uniform
interpolation grid in the BZ, for a fixed $10\times 10\times 10$ {\it ab
initio} grid. For $N=20$ the orbital magnetization per atom is already
reasonably well-converged (to within 0.002~$\mu_B$) in the case of Fe
and Co, while Ni requires $N=50$ to reach a similar level of
convergence.  Setting $N=100$ allows to converge $\M$ to better than
0.0002~$\mu_B$/atom across the board.  By comparison, the calculation of
the AHC converges much more slowly.\cite{Wang06} With $N=25$, for
example, the AHC of bcc Fe is $\sigma_{xy}=554$~S/cm, about 73\% of the
converged value of 756~S/cm, which demands a nominal mesh of the order
of $N=200$, adaptively-refined around the strongest Berry curvature
spikes.~\cite{Yao04,Wang06}
\begin{table}
\caption{\label{tab:convergence}Convergence of the orbital magnetization
of bcc Fe, hcp Co, and fcc Ni (in units of $\mu_B$/atom)
with respect to the interpolation mesh in the Brillouin zone,
for a fixed $10\times 10\times 10$ {\it ab initio} 
mesh.  For each material the magnetization is along the easy axis 
(see Table~\ref{tab:results} below).
}
\begin{tabular*}{\columnwidth}{@{}l@{\extracolsep{\fill}}ccc@{}}\hline\hline
Interpolation mesh        & bcc Fe  & hcp Co  & fcc Ni  \\ \hline
10$\times$10$\times$10    & 0.0769 & 0.0900 & 0.0461 \\
15$\times$15$\times$15    & 0.0797 & 0.0839 & 0.0394 \\
20$\times$20$\times$20    & 0.0731 & 0.0830 & 0.0455 \\
25$\times$25$\times$25    & 0.0748 & 0.0827 & 0.0535 \\ 
50$\times$50$\times$50    & 0.0749 & 0.0840 & 0.0462 \\
75$\times$75$\times$75    & 0.0760 & 0.0841 & 0.0472 \\
100$\times$100$\times$100 & 0.0761 & 0.0838 & 0.0466 \\
125$\times$125$\times$125 & 0.0760 & 0.0839 & 0.0468 \\
150$\times$150$\times$150 & 0.0760 & 0.0840 & 0.0468 \\\hline\hline
\end{tabular*}
\end{table}

Next we look at the convergence properties with respect to the {\it ab
initio} mesh, keeping the interpolation mesh fixed at $N=125$
(Table~\ref{tab:qconvergence}).  The situation is now reversed, with the
orbital magnetization converging relatively slowly compared to the
exponentially fast convergence reported in Ref.~\onlinecite{Wang06} for
the AHC.  The term $\wt{M}_z^{\rm IC}$ actually converges very rapidly,
like the AHC, but the convergence rate of $M_z$ is held back by the
larger term $\wt{M}_z^{\rm LC}$.

\begin{table}
  \caption{\label{tab:qconvergence}Convergence of the orbital
  magnetization $M_z$ (in units of $\mu_B$/atom) of bcc Fe  with respect
  to the {\it ab initio} mesh, for a fixed 125$\times$125$\times$125
  interpolation mesh. The two gauge-invariant contributions to
  $M_z=\wt{M}_z^{\rm LC}+ \wt{M}_z^{\rm IC}$ are also shown.}
\begin{tabular*}{\columnwidth}{@{}l@{\extracolsep{\fill}}ccc@{}}\hline\hline
{\it Ab initio} mesh   & Total $M_z$  & $\wt{M}_z^{\rm LC}$ & $\wt{M}_z^{\rm IC}$ \\\hline
4$\times$4$\times$4            & 0.0855 & 0.1050 & $-0.0195$ \\ 
6$\times$6$\times$6            & 0.0784 & 0.0970 & $-0.0186$ \\
8$\times$8$\times$8            & 0.0765 & 0.0948 & $-0.0183$ \\
10$\times$10$\times$10         & 0.0760 & 0.0943 & $-0.0183$ \\
12$\times$12$\times$12         & 0.0760 & 0.0943 & $-0.0183$ \\
\hline\hline
\end{tabular*}
\end{table}

In order to shed light on this behavior, we show in Table~\ref{tab:J}
the breakdown of $\wt{M}_z^{\rm LC}$ and $\wt{M}_z^{\rm LC}$, calculated
from \eqs{ic-final}{lc-final}, into the three types of terms introduced
in \equ{J0J1J2}.  The $J2$ terms give by far the largest contribution to
$\wt{M}_z^{\rm IC}(\k)$, similar to what was found previously for the
Berry curvature.~\cite{Wang06} This is not, however, the case for
$\wt{M}_z^{\rm LC}(\k)$, where the $J0$ and $J1$ terms make comparable
contributions,\cite{footnote3} and these terms are the ones limiting the
convergence rate. The reason is that they depend on matrix elements
involving the position operator, Eqs.~(\ref{eq:A})--(\ref{eq:C}). In our
implementation such matrix elements are evaluated on the {\it ab initio}
grid using the finite-differences expression (\ref{eq:disc}), and this
introduces a discretization error which decreases slowly with the grid
spacing $b$. Instead, the $J2$ terms depend exclusively on the
Hamiltonian matrix elements (\ref{eq:H}), whose convergence rate is only
limited by the decay properties of the WFs in real space (it is
therefore exponentially fast). It should be possible to achieve an
exponential convergence for the matrix elements
(\ref{eq:A})--(\ref{eq:C}) by evaluating them directly on a real-space
grid, but we have not explored that possibility in our calculations.

\begin{table}
  \caption{\label{tab:J} Decomposition of the 10$\times$10$\times$10 row
  of Table~\ref{tab:qconvergence} into the three types of terms
  appearing in \eqs{ic-final}{lc-final}, classified according to the
  number of occurrences of the matrices $J_\alpha^\pm$ [\equ{J-pm}].}
\begin{ruledtabular}
\begin{tabular}{lrrr}
& \multicolumn{1}{c}{$J0$} & \multicolumn{1}{c}{$J1$} & \multicolumn{1}{c}{$J2$} \\
\hline
$\wt{M}_z^{\rm LC}$ &    0.0397 & 0.0250 &    0.0296\\
$\wt{M}_z^{\rm IC}$ & $-$0.0002 & 0.0023 & $-$0.0204\\
\end{tabular}
\end{ruledtabular}
\end{table}

Based on the results of the convergence tests presented here, we
ultimately chose to work with a $10\times 10\times 10$ {\it ab initio}
grid and a $100\times 100\times 100$ interpolation grid for all the
calculations presented in the following section.  This choice of
parameters ensures that the values reported for the orbital
magnetization are converged with respect to $\k$-point sampling to within
0.0002~$\mu_B$/atom.

\subsection{Orbital magnetization of Fe, Co, and Ni}
\label{sec:final-results}

For each of the three materials, two separate sets of calculations were
carried out, one with the spin magnetization pointing along the easy
axis and another with the magnetization constrained to point along a
different high-symmetry direction. In each case the calculated orbital
magnetization was found to be parallel to the spin magnetization, as
expected from symmetry.

The numerical results are summarized in Table~\ref{tab:results}, where
they are compared with measurements and previous calculations.  In view
of the uncertainties in the accuracy of ordinary GGA functionals for
describing orbital-current effects, as mentioned at the end of
Sec.~\ref{sec:computational}, the overall agreement with experiment is
quite reasonable.  It can be seen that calculations based on \equ{M_def}
(both ours and those of Ref.~\onlinecite{Ceresoli_10b}) give the
ordering fcc~Ni $<$ bcc~Fe $<$ hcp~Co for the orbital magnetization per
atom, in agreement with experiment.\cite{Meyer61} Instead, the
approximate muffin-tin scheme switches the first two, because of a large
contribution in bcc Fe coming from the interstitial regions between the
muffin-tin spheres.\cite{Ceresoli_10b} The calculated anisotropy
(orientation dependence) of $\M$ is very small, and agrees reasonably
well with the one obtained in Ref.~\onlinecite{Ceresoli_10b}.

While they agree in the general trends, some difference can be seen
between the values of $\M$ obtained from \equ{M_def} of this work and in
Ref.~\onlinecite{Ceresoli_10b}. Those differences can probably be
attributed to a combination of several technical factors, including
differences in pseudopotentials, $\k$-point sampling, and our neglect of
the core-reconstruction corrections.

Regarding the two gauge-invariant contributions to $\M$ in
\equ{M-decomp}, we find that they have opposite signs in bcc Fe and hcp
Co, and the same sign in fcc Ni.  In bcc Fe and hcp Co
$|\widetilde{\M}^{\rm LC}|$ is larger than $|\widetilde{\M}^{\rm IC}|$
by a factor of about 5, while in fcc Ni that factor is more than 15.

\begin{table}
  \caption{\label{tab:results}
    Orbital magnetization (in units of $\mu_B$/atom) for bulk Fe, Co, and
    Ni.  Experimental results for $\M$ along the easy axis were obtained in
    Ref.~\onlinecite{Meyer61} by combining saturation-magnetization and
    gyromagnetic measurements.  Results from the Wannier interpolation of
    \equ{M_def} (``This work''; $\widetilde{\M}^{\rm LC}$ contributions in
    parentheses) are given and compared with the results obtained in
    Ref.~\onlinecite{Ceresoli_10b} (see footnote \onlinecite{footnote4}) by
    evaluating \equ{M_def} without Wannier interpolation (``Modern theory'')
    and by integrating currents inside muffin-tin spheres (``Muffin-tin'').
    All calculations were done using the PBE functional.}
\begin{tabular*}{\columnwidth}{@{}l@{\extracolsep{\fill}}lcccc@{}}
\hline\hline
        &       &  &\multicolumn{2}{c}{Modern theory}& Muffin-tin \\
        & Axis  & Expt.  & This work & Ref.~\onlinecite{Ceresoli_10b} & Ref.~\onlinecite{Ceresoli_10b}\\
\hline
bcc Fe  & [001]*& 0.081 & 0.0761 (0.0943) & 0.0658 & 0.0433 \\
bcc Fe  & [111] & ---   & 0.0759 (0.0944) & 0.0660 & 0.0444   \\
hcp Co  & [001]*& 0.133 & 0.0838 (0.1027) & 0.0957 & 0.0868   \\
hcp Co  & [100] & ---   & 0.0829 (0.0999) & 0.0867 & 0.0799   \\
fcc Ni  & [111]*& 0.053 & 0.0467 (0.0443) & 0.0519 & 0.0511   \\
fcc Ni  & [001] & ---   & 0.0469 (0.0440) & 0.0556 & 0.0409   \\\hline\hline
\multicolumn{5}{@{}l}{$^*$Experimental easy axis.}
\end{tabular*}
\end{table}

\subsection{Distribution of orbital magnetization in $\k$-space}

In order to understand in more detail the results of
Sec.~\ref{sec:convergence} for the convergence of the orbital
magnetization, let us look at its distribution in $\k$-space, and compare
with the AHC. For the orbital magnetization we sum the integrands in
\eqs{LC-def}{IC-def},
\beq
\label{eq:M-k}
M_z(\k)=-2\,\Im[G_{xy}(\k)+H_{xy}(\k)-2\fermi F_{xy}(\k)]\,,
\eeq
and for the AHC we take the integrand in \equ{ahc-def}, i.e., the Berry
curvature
\beq
\label{eq:curv-k}
\Omega_z(\k)=-2\,\Im\, F_{xy}(\k)\,.
\eeq
We will examine bcc Fe with the magnetization along the easy axis [001],
and accordingly we have picked the $z$-components of the axial vectors
$\M(\k)$ and ${\boldsymbol\Omega}(\k)$.

\begin{figure}
\includegraphics[width=\columnwidth]{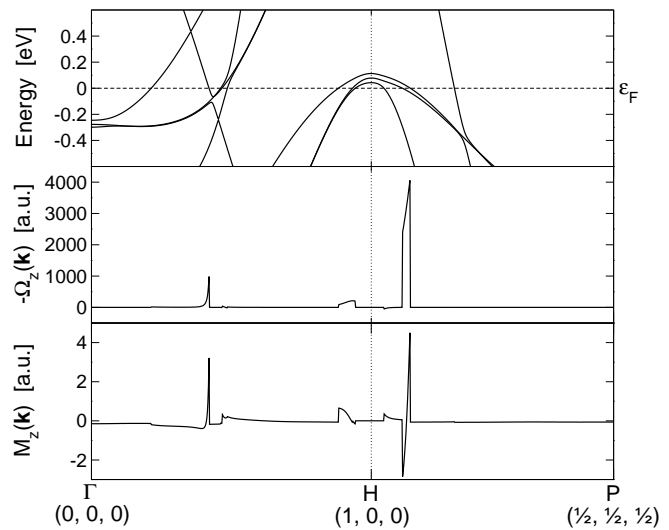}
\caption{\label{fig:fe_morb_plot} Band structure, Berry curvature
[\equ{curv-k}], and integrand of the orbital magnetization [\equ{M-k}]
calculated by Wannier interpolation along the path $\Gamma$--H--P in the
Brillouin zone. Atomic units (a.u.) are used in the middle and lower
panels.}
\end{figure}

The two quantities are plotted in Fig.~\ref{fig:fe_morb_plot} along the
high-symmetry lines $\Gamma$--H--P, together with the energy bands close
to the Fermi level. The Berry curvature (middle panel) is notorious for
its very sharp peaks, which occur when two bands, one occupied the other
empty, almost touch.~\cite{Yao04,Wang06} It can be seen in the lower
panel that $M_z(\k)$ displays similar sharp features around the same
locations, but not nearly as pronounced. The reason is that while
$\Im\,G_{xy}(\k)$ and $\Im\,H_{xy}(\k)$ are individually as spiky as
$\Omega_z(\k)$, a large degree of cancellation occurs when the three
quantities are assembled in \equ{M-k}.  This explains why the
calculation of the orbital magnetization is less demanding in terms of
BZ sampling than the AHC.

\begin{figure}
\includegraphics[width=\columnwidth]{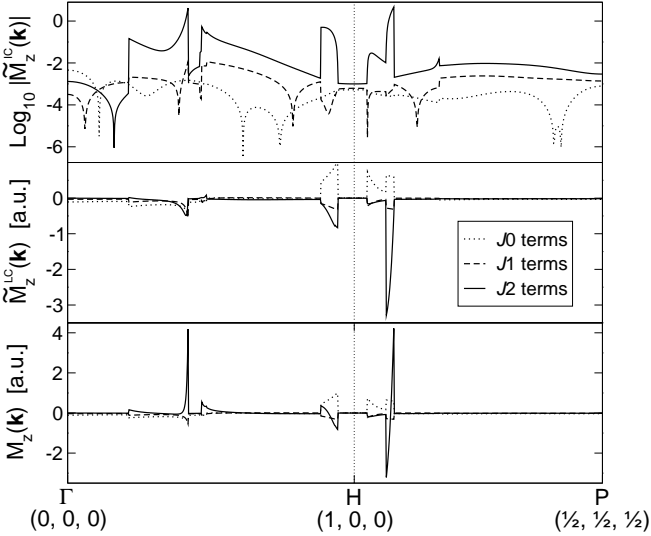}
\caption{\label{fig:fe_orb_decomp}Decomposition of the quantity
  $M_z(\k)=\widetilde{M}^{\rm LC}_z(\k)+\widetilde{M}^{\rm IC}_z(\k)$
  plotted in the lower panel of Fig.~\ref{fig:fe_morb_plot} into
  contributions from the three types of terms defined in
  Table~\ref{tab:J}.}
\end{figure}

In Fig.~\ref{fig:fe_orb_decomp} we break down the $M_z(\k)$ curve of
Fig.~\ref{fig:fe_morb_plot} into various parts.  The upper panel shows
the contributions from \equ{ic-final}, i.e.,
\beq
\wt{M}_z^{\rm IC}(\k)=-2\,\Im[H_{xy}(\k)-\fermi F_{xy}(\k)]\;,
\eeq
the middle panel those from \equ{lc-final}, i.e.,
\beq
\wt{M}_z^{\rm LC}(\k)=-2\,\Im[G_{xy}(\k)-\fermi F_{xy}(\k)]\;,
\eeq
and the lower panel their sum $M_z(\k)$. Each panel contains three
curves, labeled $J0$, $J1$, and $J2$ according to the notation of
\equ{J0J1J2} and Table~\ref{tab:J}. The $J2$ curves are the most spiky,
giving rise to the sharpest features in $M_z(\k)$. This is because
the matrices $J_\alpha^\pm$ appear twice in those terms, making them
very sensitive to small energy denominators in \equ{Jminus-fourier}. The
main features we encountered in Table~\ref{tab:J} for the integrated
quantities can already be seen in this figure. In particular, the
predominance of the $J2$ terms in $\wt{M}_z^{\rm IC}(\k)$ (note the
logarithmic scale on the upper panel of Fig.~\ref{fig:fe_orb_decomp}),
compared to a much more even distribution of $\wt{M}_z^{\rm LC}(\k)$
among the three types of terms (middle panel).

\section{Conclusions} 
\label{sec:conclusions}

We have presented a first-principles scheme, based on partially occupied
Wannier functions, to efficiently calculate the orbital magnetization of
metals using the formally correct definition for periodic crystals,
\equ{M_def}.  The localization of the WFs in real space is exploited to
carry out the necessary Brillouin-zone integrals by Wannier
interpolation, starting from the real-space matrix elements of a small
set of operators [Eqs.~(\ref{eq:HH-R}) and
(\ref{eq:AA-R}--\ref{eq:CC-R})]. The same type of strategy has
previously been used to evaluate other quantities, e.g., the anomalous
Hall conductivity~\cite{Wang06} and the electron-phonon coupling matrix
elements,~\cite{Giustino_07} which are notoriously difficult to converge
with respect to $\k$-point sampling.

As a first application, we used the method to calculate the spontaneous
orbital magnetization of the bulk ferromagnetic transition metals.
Compared to the AHC in these systems, we find that the orbital
magnetization, while displaying similar spiky features in $\k$-space
around the Fermi surface, is somewhat less demanding.  Nevertheless,
well-converged results still require a fairly dense sampling of the BZ,
making it advantageous to use an accurate interpolation scheme instead
of a direct first-principles calculation for every integration point.

Besides being computationally very efficient, the Wannier interpolation
approach has the appealing feature that the evaluation of \equ{M_def} is
done outside the first-principles code in a post-processing step. The
algorithm is in fact completely independent of such details as the basis
set used in the first-principles calculation.  As a result, the
calculation of the orbital magnetization only needs to be coded once in
the Wannier package, and the same code can be reused by interfacing it
with any desired $\k$-space electronic structure code.

We envision several possible applications for the method developed in
this work.  One example is the study of enhanced orbital moments in
low-dimensional systems, such as magnetic nanowires deposited on metal
surfaces.\cite{gambardella-nature02,komelj-prb02} The method is not
restricted to the spontaneous orbital magnetization in ferromagnets;
changes in magnetization induced by perturbations that preserve the
lattice periodicity can also be treated within the same framework. One
such application is the determination of the NMR shielding tensors
using the so-called ``converse'' approach.\cite{Thonhauser_09a,
  Thonhauser_09b} At present the converse approach, using \equ{M_def}
for the induced orbital magnetization, has only been applied to
molecules and bulk insulators. However, by combining it with the
present Wannier-based formulation, it could provide a practical route
for the evaluation of the shielding tensors in metals, a problem which
is known to demand a very dense sampling of the Brillouin
zone.\cite{davezac-prb07}

\begin{acknowledgments}
This work has been supported by NSF Grants  
DMR-10-05838 (DV) and DMR-07-06493 (IS).
We acknowledge Xinjie Wang for fruitful discussions during the early
stages of this project, and Davide Ceresoli for providing details on
the calculations reported in Ref.~\onlinecite{Ceresoli06}, including
access to the computer code. All calculations were performed on the
DEAC cluster of Wake Forest University.
\end{acknowledgments}

\appendix\section{Derivation of \equ{identity}}
\label{appendix}

With the help of \equ{f}, the left-hand-side of \equ{identity} can be
written as
\beq
\label{eq:app1}
\tr\left[ f^\hh\left( U^\dagger \AA_\alpha U\right) f^\hh
\left( U^\dagger \BB_\beta U\right)\right]\,.
\eeq
Next we use the identities
\beq
\label{eq:ident-AA}
U^\dagger \AA_\alpha U=\AA_\alpha^\hh-J_\alpha^\hh
\eeq
and
\beq
U^\dagger \BB_\beta U=\BB_\beta^\hh-\HH^\hh J_\beta^\hh
\eeq
(these can be obtained by inserting \equ{u-H} into the definitions
(\ref{eq:AA}) and (\ref{eq:BB})), to recast \equ{app1} as
\beq
\label{eq:lhs}
\sum_{nm}^J\,f_n^\hh\left(\AA^\hh_{\alpha,nm}-J^\hh_{\alpha,nm}\right)
f^\hh_m
\left(\BB^\hh_{\beta,mn}-\overline{\varepsilon}_m J^\hh_{\beta,mn}\right).
\eeq
Now we note that
\beq
\label{eq:B_occ}
f^\hh_m \BB^\hh_{\beta,mn}=
if^\hh_m\braket{u^\hh_m}{\H}{\partial_\beta u^\hh_n}=
\overline{\varepsilon}_m\AA^\hh_{\beta,mn}\,,
\eeq
since by construction the {\it occupied} Hamiltonian-gauge states are
eigenstates of the Hamiltonian (see Sec.~\ref{sec:wannier-space}). Using
\equ{B_occ} in the expression (\ref{eq:lhs}), factoring out
$\overline{\varepsilon}_m$, and invoking \equ{ident-AA} once more,
yields
\beq
\tr\left[ f^\hh\left( U^\dagger \AA_\alpha U\right) f^\hh 
\HH^\hh
\left( U^\dagger \AA_\beta U\right)\right]\,.
\eeq
Inserting $1=U^\dagger U$ between $f^\hh$ and $\HH^\hh$, the
right-hand-side of \equ{identity} is finally obtained.



\begin{thebibliography}{99}
\bibitem{footnote1} In a metal, $f_\alpha=\partial_\alpha f$ would also contain
  a $\delta$-function contribution on the Fermi surface arising from the
  discontinuity in $f$.  This is omitted here, since it is clear from
  \equ{M_def} that no special Fermi-surface contribution enters the final
  formula for $\bf M$.
\bibitem{footnote2} A decomposition similar to that of \equ{J0J1J2}
can be made for the Berry curvature, \equ{imf-b}. The $J$0,
$J$1, and $J$2 terms then correspond respectively to the
``$\overline{\Omega}$'', ``$D$-$\overline{A}$,'' and ``$D$-$D$''
terms of Ref.~\onlinecite{Wang06}.
\bibitem{footnote3} The importance of the $J0$ terms in $\wt{\M}^{\rm
  LC}(\k)$ is not surprising, given that the Wannier matrix elements of
  $\hat{\r}\H\hat{\r}$, which determine the full orbital magnetization
  in the limit of an isolated molecule, only enter through those terms.
\end{thebibliography}
\end{document}